\documentclass[twocolumn,citeautoscript,showpacs,footinbib,amsmath,amssymb,superscriptaddress,longbibliography]{revtex4-2}
\usepackage{array}
\usepackage{graphicx}
\usepackage{dcolumn}

\usepackage{xcolor}

\usepackage{bm}

\usepackage{natbib}

\usepackage{hyperref}
\usepackage{amsmath,amssymb,amsfonts,bbold}
\usepackage[normalem]{ulem}
\usepackage{float}

\usepackage{tikz}

\newcommand{\beq}{\begin{equation}}
	\newcommand{\eeq}{\end{equation}}
\newcommand{\beqa}{\begin{eqnarray}}
	\newcommand{\eeqa}{\end{eqnarray}}

\newcommand{\figEref}[1]{Extended Data Fig.\,\ref{#1}}  

\makeatletter

\newcommand{\equalcontrib}{%
	\frontmatter@footnote{These authors contributed equally to this work.}%
}

\def\frontmatter@thefootnote{%
	\ifcase\csname c@\@mpfn\endcsname
	\or \ensuremath{\dagger}%
	\or *%
	\else
	\@fnsymbol{\csname c@\@mpfn\endcsname}%
	\fi
}

\def\frontmatter@footnotemark#1{%
	\leavevmode
	\ifhmode
	\edef\@x@sf{\the\spacefactor}\nobreak
	\fi
	\begingroup
	\csname c@\@mpfn\endcsname#1\relax
	\def\@thefnmark{\frontmatter@thefootnote}%
	\@makefnmark
	\endgroup
	\ifhmode
	\spacefactor\@x@sf
	\fi
	\relax
}

\makeatother

\begin{document}	
\title{Surface-extended Andreev bound states emerging from intrinsic surface superconductivity of PtBi$_2$}
	
\author{Xiaochun Huang\equalcontrib}
\email[Corresponding author:]{xiaochun.huang@uni-wuerzburg.de}
\affiliation{Physikalisches Institut, Experimentelle Physik 2, 
	Universit\"{a}t W\"{u}rzburg, Am Hubland, 97074 W\"{u}rzburg, Germany}

\author{Lingxiao Zhao\equalcontrib}
\affiliation{Quantum Science Center of Guangdong-Hong Kong-Macao Greater Bay Area (Guangdong), Shenzhen, China}

\author{Sebastian Schimmel\equalcontrib}
\affiliation{Fakult\"{a}t f\"{u}r Mathematik und Naturwissenschaften, Bergische Universit\"{a}t Wuppertal, Wuppertal, Germany}	 
\affiliation{Leibniz-Institute for Solid State and Materials Research (IFW-Dresden), Dresden, Germany}	

\author{Julia Besproswanny}
\affiliation{Fakult\"{a}t f\"{u}r Mathematik und Naturwissenschaften, Bergische Universit\"{a}t Wuppertal, Wuppertal, Germany}	
\affiliation{Leibniz-Institute for Solid State and Materials Research (IFW-Dresden), Dresden, Germany}	

\author{Luminita\,Harnagea}
\affiliation{Leibniz-Institute for Solid State and Materials Research (IFW-Dresden), Dresden, Germany}	
\affiliation{I-HUB Quantum Technology Foundation, Indian Institute of Science Education and Research, Pune 411008, India}

\author{Patrick H\"{a}rtl} 
\affiliation{Physikalisches Institut, Experimentelle Physik 2, 
	Universit\"{a}t W\"{u}rzburg, Am Hubland, 97074 W\"{u}rzburg, Germany}	

\author{Christian Hess}
\affiliation{Fakult\"{a}t f\"{u}r Mathematik und Naturwissenschaften, Bergische Universit\"{a}t Wuppertal, Wuppertal, Germany}	 	
\affiliation{Leibniz-Institute for Solid State and Materials Research (IFW-Dresden), Dresden, Germany}

\author{Bernd B\"{u}chner}
\affiliation{Leibniz-Institute for Solid State and Materials Research (IFW-Dresden), Dresden, Germany}	 
\affiliation{Technische Universit\"{a}t Dresden, Dresden, Germany}

\author{Matthias Bode} 
\affiliation{Physikalisches Institut, Experimentelle Physik 2, 
	Universit\"{a}t W\"{u}rzburg, Am Hubland, 97074 W\"{u}rzburg, Germany}
\affiliation{Wilhelm Conrad R{\"o}ntgen-Center for Complex Material Systems (RCCM), Universit\"{a}t W\"{u}rzburg, Am Hubland, 97074 W\"{u}rzburg, Germany} 	


	
	
\date{\today}
	
\begin{abstract}
Intrinsic topological superconductivity remains one of the central open questions in condensed-matter physics. The three-dimensional Weyl semimetal PtBi$_2$ {has recently been shown} by angle-resolved photoemission spectroscopy to host a superconducting gap that opens exclusively on its Fermi-arc surface states and exhibits a nodal structure, {making it a promising platform for exploring unconventional and potentially topological superconductivity}. However, despite these momentum-resolved {insights, the real-space spectroscopic signatures of unconventional superconductivity in PtBi$_2$ remain unclear}. Here, using scanning tunneling microscopy and spectroscopy, {we reveal a sizable and spatially homogeneous gap ($\Delta > 10\,\mathrm{meV}$) on atomically pristine terraces of PtBi$_2$}. Within this gap, we {detect surface-extended in-gap states exhibiting pronounced atomic-scale spatial modulation}. {Analysis of the tunneling spectra indicates that these states are most consistently interpreted as Andreev bound states within a superconducting gap. Further phenomenological modeling shows that their characteristic spectral features are naturally captured by a chiral pairing scenario. Our results establish a direct real-space connection between the superconducting gap and surface-extended quasiparticle states in PtBi$_2$, and provide new insight into the nature of its unconventional surface superconductivity. These findings highlight PtBi$_2$ as a versatile platform for exploring exotic quasiparticle excitations and potential Majorana physics.}
\end{abstract}
	
\maketitle
	
	
\section{I\MakeLowercase{ntroduction}}   
\label{sec:Introduction}
\vspace{-0.3cm}

Topological superconductors offer a fertile platform for realizing Majorana zero modes (MZMs)---topologically protected, zero-energy quasiparticles that are robust against local perturbations and hold promise for fault-tolerant quantum computing \cite{R1,R2,R3}. According to topological band theory and the Bogoliubov-de Gennes {(BdG)} formalism, such a phase can emerge when superconductivity opens a gap in a spin-momentum-locked electronic band \cite{R1,R2}. Guided by this principle, extensive efforts have pursued engineered artificial architectures, commonly referred to as effective $p$-wave superconductors, by inducing pairing in spin-momentum-locked states. Using scanning tunneling microscopy/spectroscopy (STM/STS), signatures of topological superconductivity have been reported at the ends of magnetic atom chains on conventional superconductors \cite{R7,R8}, on individual Fe impurities at the surface of Fe(Te,Se) \cite{J.Yin_Fe}, and inside vortex cores of FeTe$_{0.55}$Se$_{0.45}$ single crystals and Bi$_2$Se$_3$/Bi$_2$Te$_3$--NbSe$_2$ heterostructures \cite{R11,R12,R8,R9}. 

Despite these advances, major challenges persist: the need for complex fabrication, reliance on magnetic fields, and---most critically---their signals are tied to local microstructures. A promising alternative is to identify intrinsic topological superconductors, where native surface states provide a clean spectroscopic window into topological superconductivity. However, whether such materials truly exist remains under debate \cite{R10,R14,R15}. Even in proposed candidates, the superconducting (SC) gaps are extremely small---typically requiring millikelvin temperatures for sufficient energy resolution---which has so far hindered definitive spectroscopic verification of topological superconductivity \cite{R15}.

\begin{figure*}[t] 
	\centering
	\includegraphics[width=0.95\textwidth]{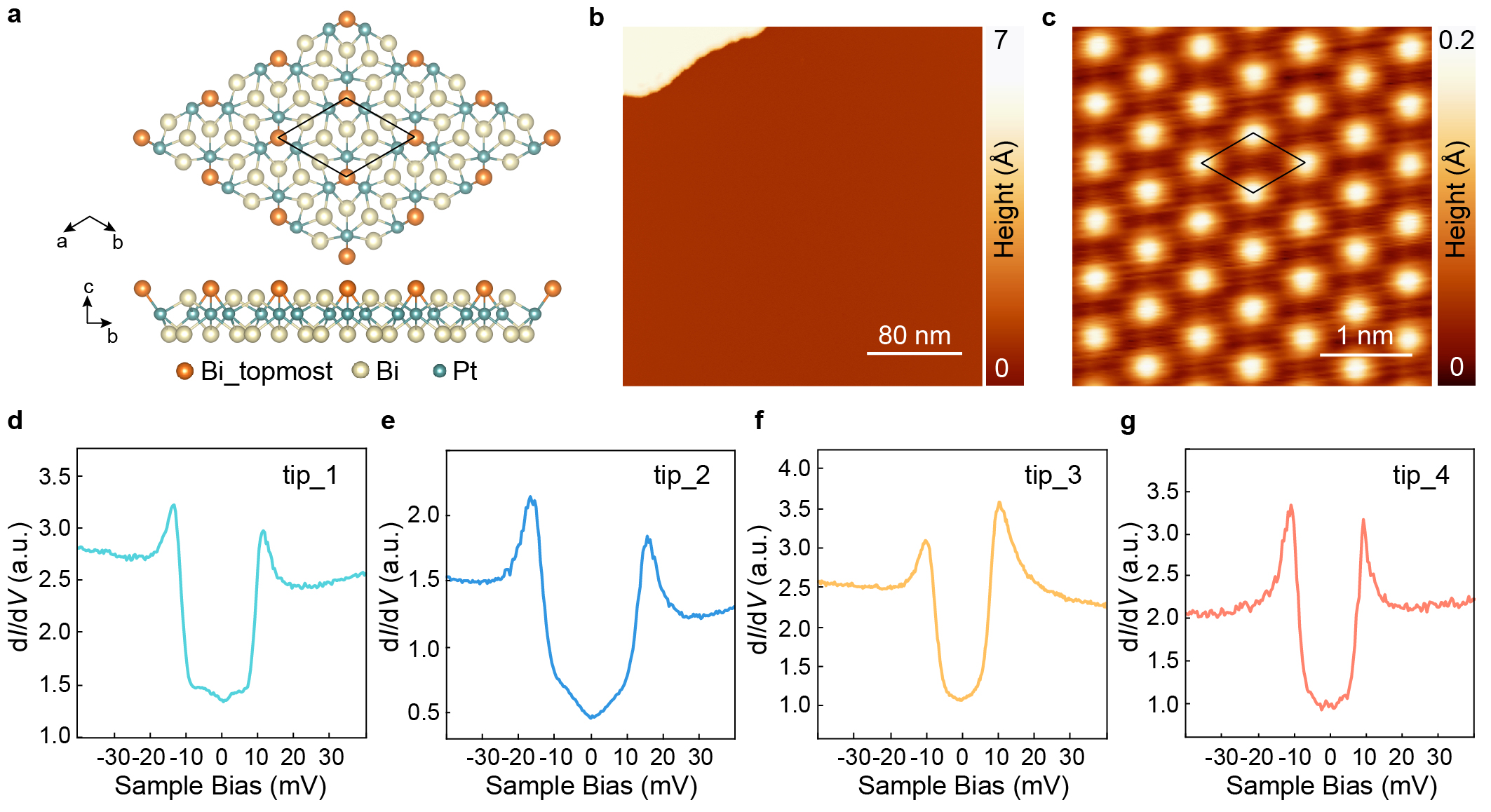}
	\caption{\textbf{{Sizable gap} on the decorated honeycomb surface of trigonal PtBi$_2$.} 
		\textbf{a}, Schematic structural model of a monolayer PtBi$_2$.
		Top panel: Top view of the decorated honeycomb surface, with the unit cell outlined by a diamond. 
		Bottom panel: Side view, highlighting the topmost Bi atoms (shown in orange in the top panel). 
		\textbf{b}, Large-scale STM image of a cleaved PtBi$_2$ surface ($V = -2.0$\,V, $I = 10$\,pA). 
		\textbf{c}, Atomic-resolution STM image ($V = -50$\,mV, $I = 1.0$\,nA), showing a hexagonal arrangement 
		of the topmost Bi atoms on the decorated honeycomb surface. The diamond marks a unit cell. 
		\textbf{d}-\textbf{g}, Representative d$I$/d$V$ spectra acquired with four different W tips on the same sample. 
		The tips are denoted as tip\_1, tip\_2, tip\_3, and tip\_4, and this designation is used consistently throughout the manuscript. 
		Measurement conditions: $V_{\text{stab}} = 40$\,mV, $I_{\text{stab}} = 200$\,pA
		for \textbf{d}, \textbf{e}, and \textbf{g}; 
		$V_{\text{stab}} = 40$\,mV, $I_{\text{stab}} = 500$\,pA for \textbf{f}. $V_{\text{mod}} = 0.5$\,mV, $T$ = 5.1\,K.
		All spectra are shown as raw data.} 
	\label{Figure1}
\end{figure*} 

Weyl semimetals, characterized by topologically protected Fermi-arc surface states and bulk Weyl nodes, have been theoretically predicted and experimentally realized \cite{R16,R17,R18,R19,TaAs_EP2}. As members of the topological materials family, their SC properties have attracted growing interest, though most studies to date have focused on bulk electronic states \cite{R20,R21,R22,R23}. Recently, high-resolution angle-resolved photoemission spectroscopy (ARPES) studies have identified trigonal PtBi$_2$ as a three-dimensional (3D) Weyl semimetal, with a SC gap opening exclusively on the Fermi arc surface states at temperatures around $10\,\mathrm{K}$ \cite{R4,iwave,Andrii_ARPES_2}, {although bulk superconductivity has not been conclusively established.} Subsequent ARPES further revealed a node in the order parameter at the center of the Fermi arc, leading to the proposal of an $i$-wave pairing symmetry ($\ell\,=\,6$) in PtBi$_2$ \cite{iwave}. 
Combined with the characteristic spin texture of Fermi arcs in Weyl semimetals \cite{R25}, 
these findings position PtBi$_2$ as a promising candidate for intrinsic topological superconductivity \cite{iwave}. 

Adding to its unconventional character, PtBi$_2$ has been found to host exceptionally large surface SC gaps \cite{R5}. 
While ARPES measurements report values of approximately 1.4 to 2.0\,meV \cite{R4,iwave}, STM/STS studies indicate that the gap can locally reach up to 20\,meV \cite{R5}. At the same time, STM/STS has also revealed elusive and spatially inhomogeneous gap features, including regions where the SC gap appears to close entirely without any apparent correlation to surface defects or structural irregularities \cite{R5,R26}. {Such inhomogeneity may disrupt the spatial coherence of the pairing potential and {hinder the emergence of novel surface modes.} These findings have sparked considerable interest in exploring potential nontrivial SC properties and in investigating the prospect of high-temperature Majorana physics in this material \cite{R24,R27,R28,R29,R31}.}
\begin{figure}[t] 
	\centering
	\includegraphics[width=0.45 \textwidth]{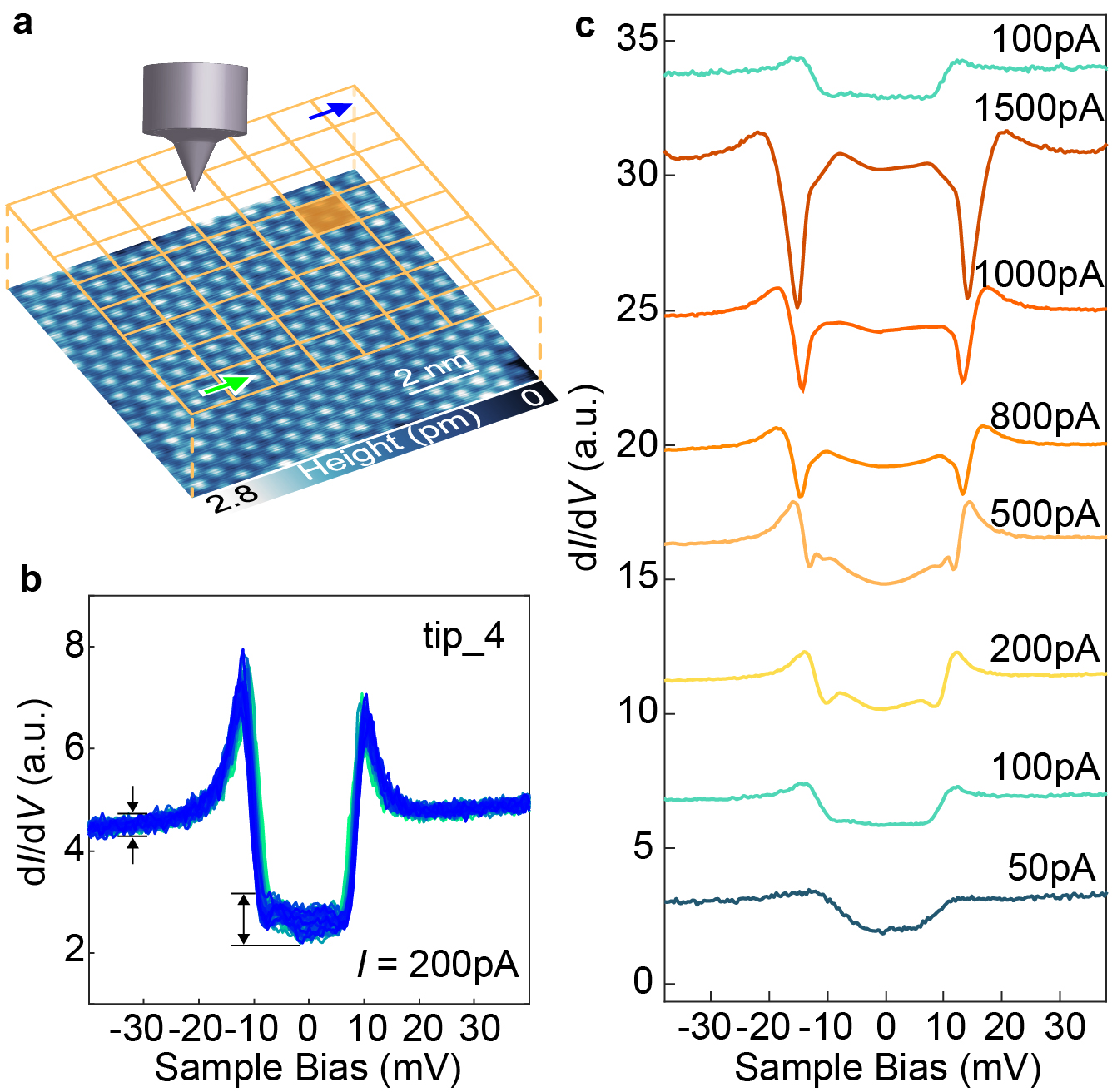}
	\caption{\textbf{Spatially {homogeneous gap} and the tip--sample distance dependence of in-gap states.} 
		\textbf{a}, Atomic-resolution STM image ($V = 200$\,mV, $I = 500$\,pA) of a ${(10 \times 10)}$ nm\textsuperscript{2} region 
		of the decorated honeycomb surface of PtBi$_2$. A yellow ${(8 \times 8)}$ grid is overlaid on the image.	
		\textbf{b}, d$I$/d$V$ spectra measured at the center of each square in the grid shown in \textbf{a}. 
		The measurements began at the leftmost point of the bottom row (green arrow) and proceed sequentially along the row. After completing each row, the tip moved to the beginning of the next row and repeated the process, 
		ending at the rightmost point of the top row (blue arrow). 	
		\textbf{c}, d$I$/d$V$ spectra acquired at the position marked by the orange square in \textbf{a}.
		Measurement conditions: $V_{\text{stab}} = 40$\,mV, $V_{\text{mod}} = 0.5$\,mV, and $T$ = 5.1\,K.
		The spectra are vertically offset for clarity. All spectra are presented as raw data.} 
	\label{Figure2}
\end{figure} 

{\section{S\MakeLowercase{izable} \MakeLowercase{gap in }P\MakeLowercase{t}B\MakeLowercase{i}$_2$}}   
\label{Sec:ExpSetupProc}
\vspace{-0.3cm}

Trigonal PtBi$_2$ is a layered material with a non-centrosymmetric space group ($P31m$), 
in which each monolayer adopts a Bi--Pt--Bi sandwich structure \cite{R32,R33}. 
Owing to the absence of inversion symmetry, cleaved PtBi$_2$ crystals exhibit two distinct surface terminations: 
A twisted kagome lattice and a decorated honeycomb lattice \cite{R29}. 
Surface superconductivity has been reported on both terminations \cite{R4,R5}. 
In this study, we focus on the decorated honeycomb surface shown in Fig.\,\ref{Figure1}a, where the topmost Bi atoms 
form a simple triangular lattice, in contrast to the more intricate structure of the twisted kagome termination \cite{R29}.

Figure\,\ref{Figure1}b shows a large-scale STM image of the freshly cleaved surface of a PtBi$_2$ single crystal. 
An atomic-resolution STM image (Fig.\,\ref{Figure1}c) reveals a triangular surface lattice, 
consistent with the decorated honeycomb termination. 
To establish whether the sample hosts
{a sizable gap consistent with surface superconductivity}, 
we conducted differential conductance (d$I$/d$V$) spectroscopy. 
As shown in Fig.\,\ref{Figure1}d-g, pronounced {SC-like gaps} with well-defined coherence peaks 
symmetrically centered around the Fermi level ($E_{\text{F}}$) are reproducibly observed using four different tips.
{Fitting the tunneling spectra with isotropic (and anisotropic) Dynes models \cite{Dynes1978} yields
gap magnitudes of 11.4\,meV (13.2\,meV), 14.0\,meV (16.1\,meV), 8.5\,meV (10.1\,meV), and 8.9\,meV (10.1\,meV), respectively (see Methods and Fig.S1 in the Supplementary Information).}
{We note that the finite density of states (DOS) within the gap prevents an unambiguous distinction between isotropic and anisotropic scenarios. The finite zero-bias conductance can arise from contributions of bulk metallic electronic states \cite{R4,R5}, as well as from low-energy quasiparticle excitations associated with the SC state (see Supplementary Information).}

Since changing the tip often introduces lateral shifts on the order of sub-millimeters, 
these results indicate that {sizable gaps} persist across such macroscopic length scales, 
although their magnitudes exhibit measurable variation. 
{Notably, we reproducibly observe similarly large gaps ($\Delta > 10\,\mathrm{meV}$) in PtBi$_2$ samples obtained from different research groups (see Supplementary Information), characterizing it as a robust and intrinsic property of PtBi$_2$.}
\begin{figure*}[t] 
	\centering
	\includegraphics[width=1.0\textwidth]{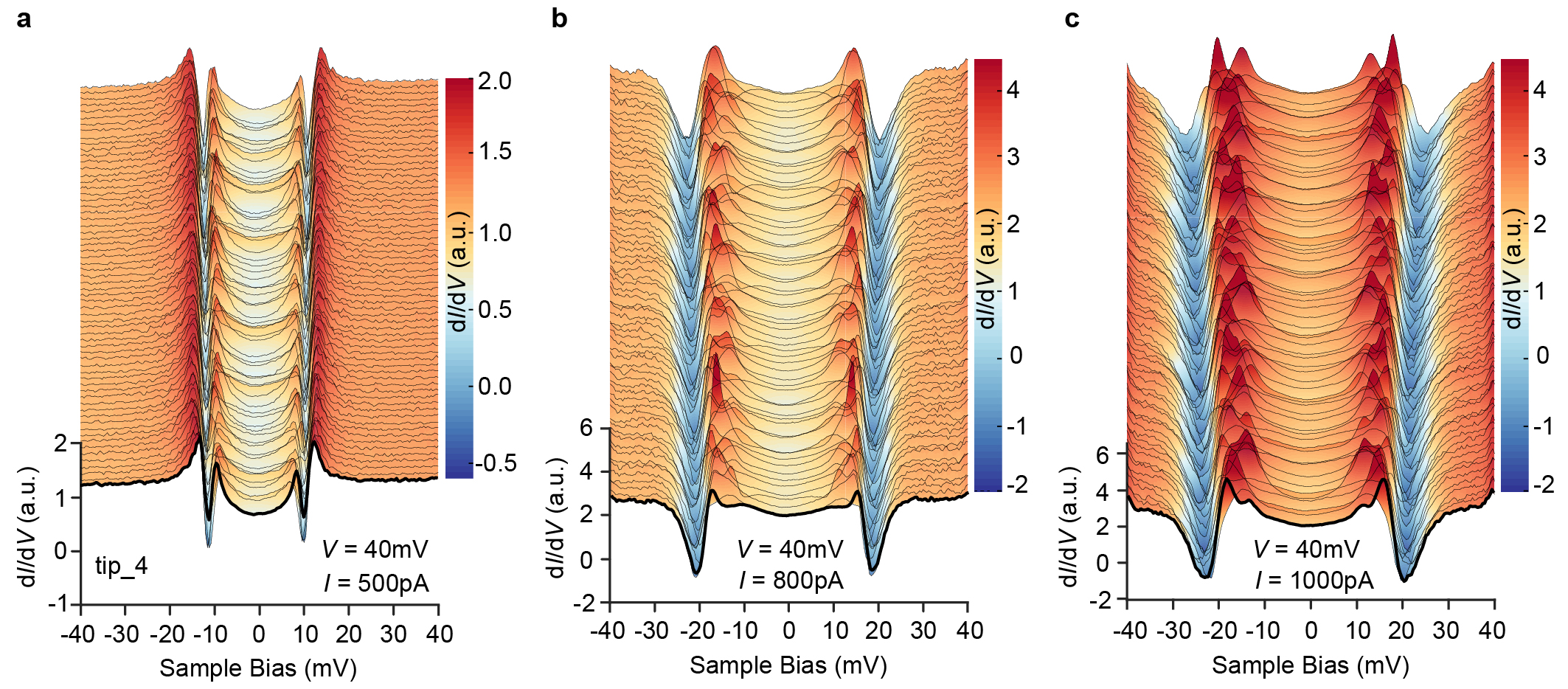}
	\caption{\textbf{Spatially extended {in-gap states}.} 
		\textbf{a}-\textbf{c}, Intensity plots overlaid with line cuts of d$I$/d$V$ spectra 
		acquired from the same region as Fig.\,\ref{Figure2}a, using the identical ${(8 \times 8)}$ grid pattern. 
		Measurement conditions: $V_{\text{stab}} = 40$\,mV, $V_{\text{mod}} = 0.5$\,mV, and $T$ = 5.1\,K. 
		The setpoint tunneling currents were 500\,pA, 800\,pA, and 1000\,pA for \textbf{a}, \textbf{b}, and \textbf{c}, respectively.
		In each panel, the first spectrum (highlighted in bold) serves as a reference and is plotted against the left axis. 
		All spectra are shown as raw data.}				 
	\label{Figure3}
\end{figure*} 

We next examine the homogeneity {of the gap over length scales ranging} from hundreds of nanometers down to the atomic scale. 
As shown in \figEref{FigureE1}, spectra acquired at positions separated by ${\sim}700 \text{ nm}$ remain highly uniform, with a standard deviation below $1\,\mathrm{meV}$.
Figure\,\ref{Figure2}a,b show spatially resolved d$I$/d$V$ spectra acquired over a {clean surface area using a 2D grid with an 1.25\,nm increment. The coherence peaks coincide at all grid points,} indicating a uniform $10\,\mathrm{mV}$ gap on the nanometer scale. Additional measurements (\figEref{FigureE2}) further demonstrate that {the gap width} remains homogeneous down to {length scales smaller than the surface lattice constant}.
{We note that the reported gap magnitude in PtBi$_2$ varies across different studies, suggesting a strong sensitivity of the SC properties to surface electronic conditions. For example, a similarly large gap has been reported in a previous STM study and was assigned to a SC origin based on its magnetic-field dependence \cite{R5}. By contrast, ARPES measurements \cite{R4,iwave,Andrii_ARPES_2} and recent STM/STS studies \cite{chenhui_CPL} have reported substantially smaller gap values. A detailed comparison of multiple independent studies is provided in the Supplementary Information. \vspace{0.5cm}}

{\section{I\MakeLowercase{n-gap state} \MakeLowercase{features}}}  
\label{Sec:Results} 
\vspace{-0.3cm}

As indicated by the arrows in Fig.\,\ref{Figure2}b, the conductance intensity within {the gap} 
exhibits stronger spatial variation than the relatively flat background outside the coherence peaks. 
This variation is more clearly visualized in \figEref{FigureE3}, where the in-gap states
display unexpected spatial inhomogeneity. Given that the atomic-resolution STM image of the measured area (Fig.\,\ref{Figure2}a) 
shows no correlation with surface defects, the observed spatial variation in the in-gap states points to an origin beyond defect-induced states. This interpretation is substantiated by the measurements shown in Fig.\,\ref{Figure2}c, where a series of d$I$/d$V$ spectra 
were acquired at the same location but by varying the stabilization tunneling current ($I_{\text{stab}}$), 
thereby probing the response of the in-gap states to changes of the tip--sample distance.  
At low $I_{\text{stab}}$ (100\,pA), the spectrum exhibits a typical {SC-like} gap with a featureless in-gap region. 
As $I_{\text{stab}}$ increases, corresponding to a reduced tip--sample distance, 
distinct in-gap states begin to rise from the flat background. 
At $I_{\text{stab}} = 800$\,pA, these states evolve into a well-defined spectral structure: 
A broad double-hump symmetrically positioned around $E_{\text{F}}$. 
Further increases in $I_{\text{stab}}$ (1000\,pA, and 1500\,pA) enhance the spectral weight of the in-gap states 
and shift the outer spectral features farther from $E_{\text{F}}$, although the overall profile remains essentially unchanged. Crucially, when $I_{\text{stab}}$ is reduced back to 100\,pA, the {original gap} is fully recovered, 
thereby confirming the reversibility of the observed evolution. 
Similar behavior is observed with multiple tips, underscoring the reproducibility 
and robustness of this phenomenon (see \figEref{FigureE4}).

To investigate the spatial characteristics of the observed {in-gap states}, 
we repeated the measurements of Figs.\,\ref{Figure2}a,b at progressively higher $I_{\text{stab}}$, 
acquiring spatially resolved d$I$/d$V$ spectra over the same grid. 
The data are visualized as 3D surface plots in Fig.\,\ref{Figure3} and as overlaid spectra in \figEref{FigureE5}. 
At $I_{\text{stab}} = 500$\,pA (Fig.\,\ref{Figure3}a and \figEref{FigureE5}a), the spectra remain highly consistent across all positions, 
exhibiting a well-defined, symmetric profile about $E_{\text{F}}$, aside from minor variations near zero bias. 
A representative spectrum (highlighted in bold in Fig.\,\ref{Figure3}a) features a conductance minimum at $E_{\text{F}}$, 
followed by a steep increase in spectral weight toward the gap edges. 
Just before the coherence peaks, the DOS develops a sharp local dip, producing two pronounced peaks 
symmetrically positioned about $E_{\text{F}}$, thereby forming a characteristic double-humped structure within {the gap}.
Beyond this dip, the spectrum rapidly recovers and connects nearly vertically to the coherence peaks. 
In this configuration, the {in-gap states} coexist with distinct coherence peaks of higher intensity, 
and the abrupt spectral transition at their interface gives rise to two conductance minima, 
which at some positions dip below zero (more clearly seen in \figEref{FigureE5}). With increasing $I_{\text{stab}}$ (Fig.\,\ref{Figure3}b,c), the spectral weight and the energy spread of the {in-gap states} 
are significantly enhanced, {progressively obscuring the gap features}.
Figure\,\ref{Figure3}a-c collectively demonstrates that these {in-gap states} are spatially extended across the surface.
This surface-extended nature is further corroborated by high-resolution measurements using a different tip, 
with sub-lattice constant spatial sampling, as shown in \figEref{FigureE6}.
 
\begin{figure}[t]
	\centering
	\includegraphics[width=0.45\textwidth]{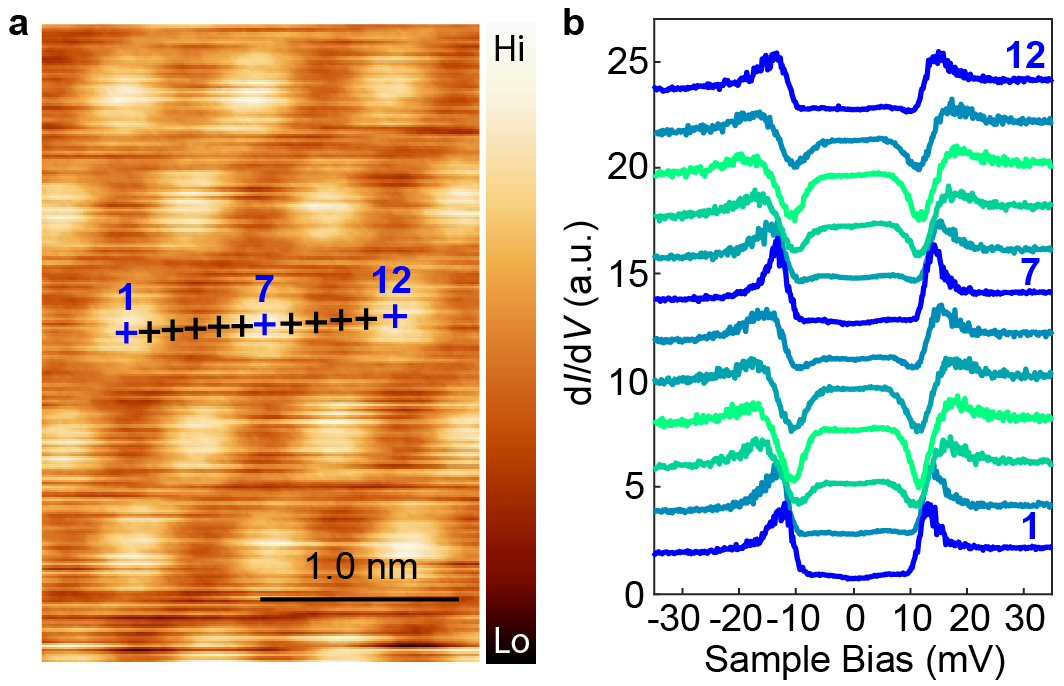}%
	\caption{\textbf{{Spatial variation of in-gap states within a single surface unit cell of PtBi$_2$.}}  
		{\textbf{a}, Atomic-resolution STM image of PtBi$_2$ ($V = 200$\,mV, $I = 200$\,pA), resolving the topmost Bi atoms arranged in a hexagonal lattice.}
		{\textbf{b}, Spatially resolved d$I$/d$V$ spectra acquired at positions indicated in \textbf{a}. Measurement conditions: $V_{\text{stab}} = 40$\,mV, $I_{\text{stab}} = 800$\,pA, and $V_{\text{mod}} = 0.5$\,mV, $T$ = 5.1\,K. Spectra are vertically offset for clarity. All spectra are presented as raw data.}
	}
	\label{Figure4} 
\end{figure}

{It is worth noting that Figs.\,\ref{Figure3}b,c reveal pronounced spatial variations in the spectra of the in-gap states. This is confirmed by higher-spatial-resolution STM/STS measurements (see Fig.\,\ref{Figure4}).
As shown, the in-gap features are strongly suppressed when spectra are acquired directly on top of the topmost Bi atoms (e.g., spectra 1, 7, and 12). Moving between neighboring Bi atoms (for example, from 1 to 7), the in-gap spectral weight increases continuously, reaches a maximum midway between the two atoms, and subsequently decreases as the tip approaches the next Bi site. The same spatial modulation repeats from 7 to 12. These variations are reproducible across different samples, demonstrating a clear atomic-scale spatial modulation of the in-gap states.
The observed modulation can be understood as arising from the interplay between the position-dependent tunneling matrix element, the momentum-dependent quasiparticle band, and the momentum-dependent SC order parameter \cite{R1,AR_PRL_2023} (as discussed below).}	
\vspace{0.5cm}

\vfill

\begin{figure*}[t] 
	\centering
	\includegraphics[width=0.85\textwidth]{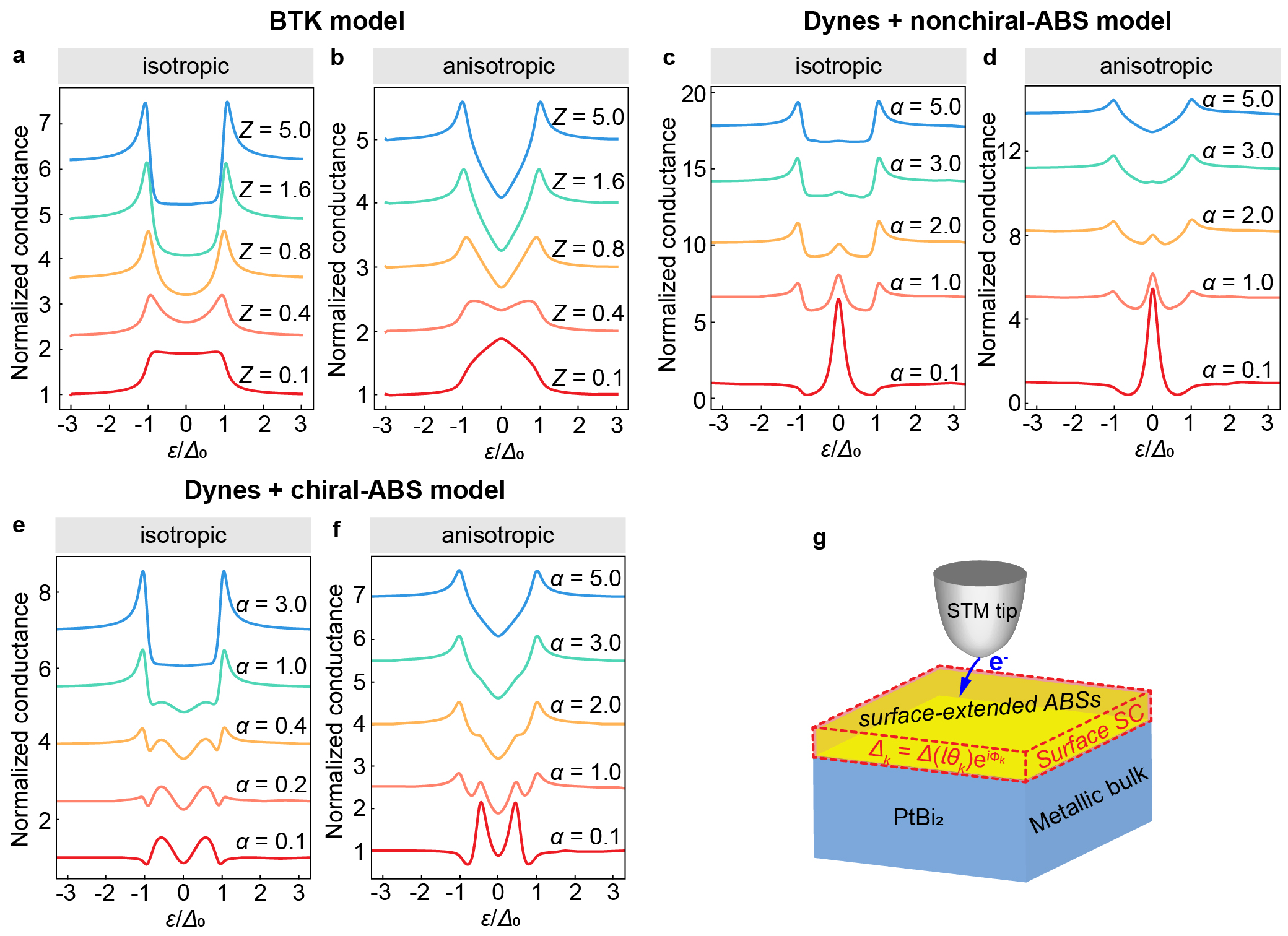}
	\caption{{\textbf{Possible superconducting origins of the observed in-gap states.} 	
		\textbf{a},\textbf{b}, Calculated d$I$/d$V$ spectra within the BTK models for isotropic \textbf{a} and anisotropic \textbf{b} pairing symmetries. $Z$ characterizes the effective tip--sample junction transparency.
		\textbf{c},\textbf{d}, Dynes $+$ nonchiral-ABS model for isotropic \textbf{c} and anisotropic \textbf{d} pairing.
		\textbf{e},\textbf{f}, Dynes $+$ chiral-ABS model for isotropic \textbf{e} and anisotropic \textbf{f} pairing. The relative weight of the Dynes and chiral-ABS contributions is controlled
		by $\alpha$. All simulations were performed at $T = 6$ K. Spectra are vertically offset for clarity.
		\textbf{g}, Schematic of STM tunneling into ABSs in PtBi$_2$. The red dashed box indicates the finite-thickness surface SC region, within which the surface-extended ABSs are highlighted in yellow. The pairing is described by $\Delta_k = \Delta(l\theta_k)\, e^{i\phi_k}$, where $\theta_k$ denotes the polar angle, the integer $l$ specifies the angular symmetry of the order parameter, and $\phi_k$ represents the momentum-dependent phase. The blue slab represents the metallic bulk of PtBi$_2$.} 
	}				 
	\label{Figure5}
\end{figure*} 

{\section{{P\MakeLowercase{hysical origin of the in-gap states}}}}   
\label{Sec:Results} 
\vspace{-0.3cm}

{We now discuss the possible physical origins of the observed spectra. These are characterized by a wide, symmetric gap, pronounced coherence peaks, and double-hump-like in-gap states that depend strongly on both the tunneling resistance and the spatial position.} 
{Two potential origins of these observations were investigated: 
(i) superconductivity-related low-energy excitations, including interfacial Andreev reflection processes and intrinsic in-gap states such as Andreev bound states (ABSs), whose spectral features can provide insight into the SC pairing symmetry and its possible unconventional nature \cite{ABS_Bruder,ABS_review, ABS_SrRuO_1,ABS_SrRuO_2, ABS_NoncentrosymmetricSC,ABS_3DTIS,R15,BTK,AR_PRL_2023}}.
{Notably, the spectral features in the present work share key qualitative similarities with those reported in Refs. \cite{ABS_SrRuO_1,ABS_SrRuO_2, ABS_NoncentrosymmetricSC,ABS_3DTIS,R15}.}

\vspace{0.1cm}
{An alternate, more conventional interpretation (ii) involves inelastic electron tunneling processes associated with H$_2$. Previous STM/STS studies on other material systems have shown that H$_2$ adsorbates can generate several spectral features resembling those observed here, including SC-gap-like structures, negative differential conductance (NDC), and double-hump-like in-gap features \cite{PhysRevB.71.115416,PhysRevLett.111.146102}. As H$_2$ is a common residual gas in ultra-high-vacuum systems, adsorption on either the tip or the sample surface must therefore be considered.} 
{To examine scenario(ii), we performed systematic control experiments (see Supplementary Information for details). The results indicate that the spectral features shown in Figs.\,\ref{Figure2}-\ref{Figure4} are neither tip-induced artifacts nor cause by H$_2$ contamination of the PtBi$_2$ sample, thereby strongly disfavoring an H$_2$-related inelastic tunneling origin.}

\vspace{0.1cm}
{To assess the remaining superconductivity-related origin (i), we employ two phenomenological models for the tunneling spectra, capturing contributions from Andreev reflection and from an ABS-based mechanism, respectively.
We first consider Andreev reflection within the Blonder-Tinkham-Klapwijk (BTK) framework \cite{BTK}:}
\begin{equation}
		{I = I_{\mathrm{BTK}}(U;Z,\Delta,T)  
		\label{eq_BTK}}
	\end{equation}
where $U$ is the applied bias voltage, corresponding to an energy $\varepsilon = eV$. The superconducting gap is introduced as $\Delta$, and $T$ denotes the temperature. $Z$ characterizes the effective tip--sample junction transparency and is therefore varied with tip--sample distance (see Methods for details).
{Using this model, we calculate tunneling spectra for different $Z$ values and for both isotropic and anisotropic pairing symmetries (Figs.\,\ref{Figure5}a,b). As shown, large $Z$ (corresponding to a large tip--sample distance) yields tunneling-dominated spectra with a well-defined SC gap, whereas decreasing $Z$ drives a continuous evolution of the line shape towards an Andreev-reflection-dominated regime, with enhanced spectral weight appearing within the gap energy range. Importantly, this evolution does not correspond to the emergence of sharp in-gap states within a persisting SC gap. Accordingly, the BTK model fails to reproduce the experimentally observed coexistence of the SC gap and the double-hump-like in-gap states, as well as the evolution of their relative spectral weight with different tip--sample distances.} 
{Note, that in our tunneling  geometry,  where the injected electrons are far away from the edges of the putative 2D SC state, the appearance of a double peak as a possible signature of a chiral order parameter \cite{R1} is not expected in the BTK model, even if such is considered \cite{ABS_Satoshi}. We therefore refrain from considering the Andreev tunneling scenario which is modelled by the BTK approach with a varying $Z$ which parametrises the Andreev scattering strength in dependence of the tip sample distance any further.}

Instead, we investigate the possibility that the observed double peak structure {is indeed} the signature of an intrinsic surface-extended ABS similar to that expected in superconducting topological insulators (STI) \cite{ABS_3DTIS}. Since the detailed tunneling conductance for a putative {ABS of a non-time-reversal-symmetry-breaking} Weyl semimetal, to the best of our knowledge, has not been worked out yet, we employ the extended BTK-type framework \cite{ABS_Satoshi_2, ABS_Satoshi} which was originally proposed for modeling the mentioned STIs \cite{ABS_3DTIS}. In this approach, {the tunneling current is written as}
\begin{equation}
	{I = I_{\mathrm{ABS}}(U;Z_\mathrm{int},\Delta,T)  
		\label{ABS_only}.}
\end{equation}
The framework allows us to calculate tunneling spectra for a chiral order parameter of the form {$\Delta_k = \Delta(l\theta_k)\, e^{i\phi_k}$} (see Methods). Here, $Z_\mathrm{int}$, distinct from the barrier parameter $Z$ in the conventional BTK model, describes the effective scattering strength at the intrinsic interface between the SC surface states and the non-SC metallic bulk states. It is therefore not expected to vary with the tip--sample distance.
Varying $Z_\mathrm{int}$ modifies the detailed spectral line shape, but does not reproduce the experimentally observed tip--sample-distance-dependent evolution (see \figEref{FigureE7}).

{In the ABS scenario, it is reasonable to assume that the tip--sample tunneling probes the superconducting condensate and the ABS with different tunneling matrix elements, which qualitatively explains the observed tip sample distance.
	In lack of a proper theory for the latter, we empirically consider a combined Dynes + ABS model.}
{Within this model, the tunneling current is written as:}
\vspace{0.1cm}
\begin{equation}
		{
		I = \left(1 - \mathrm{e}^{-\alpha}\right) \cdot I_{\mathrm{Dynes}}
		+ \mathrm{e}^{-\alpha} \cdot \, I_{\mathrm{ABS}},
		}
		\label{Dynes_ABS_model}
\end{equation}
{where $I_{\mathrm{Dynes}}$ describes the conventional tunneling contribution within the Dynes model, whereas $I_{\mathrm{ABS}}$ captures the ABS contribution. The intrinsic ABS contribution is calculated with a fixed $Z_\mathrm{int}$, while the relative weight of $I_{\mathrm{Dynes}}$ and $I_{\mathrm{ABS}}$ is controlled by the parameter $\alpha$, which varies systematically with tip--sample distance. Further details of the model are provided in the Methods.}

{Using Eq.\,\ref{Dynes_ABS_model}, the model reproduces spectra dominated by the SC gap at larger $\alpha$ and by in-gap ABS features at smaller $\alpha$ (Figs.\,\ref{Figure5}c-f). For intermediate $\alpha$, it captures the coexistence of these two spectral components, closely resembling the experimental evolution observed in Fig.\,\ref{Figure2}. The simulations further show that pairing chirality is essential to reproduce the characteristic double-hump-like in-gap features (Figs.\,\ref{Figure5}e,f), whereas nonchiral pairing yields single-peak-like in-gap features (Figs.\,\ref{Figure5}c,d). By contrast, isotropic and anisotropic gap amplitudes mainly affect the sharpness and detailed line shape of the in-gap features, but do not qualitatively alter their double-peak structure.} 
We note that pronounced negative differential conductance (NDC) features are observed in the experimental spectra (Fig.\,\ref{Figure3}). Although the Dynes $+$ ABS model does not quantitatively reproduce the NDC, it captures the corresponding spectral evolution qualitatively, yielding two pronounced dips near the edges of the in-gap states that remain, however, positive (Figs.\,\ref{Figure5}e,f). While the microscopic origin of the NDC remains unclear, such features are generally associated with a rapid variation in the effective tunneling transmissivity, particularly across the transition between spectral peaks and dips {\cite{Britnell2013,PhysRevB.97.115442}}.
	
These results indicate that {some key-aspects of the experimental data can consistently be} described by a scenario in which the observed in-gap states arise from intrinsic ABSs associated with a surface ABS band within a SC gap.
Within this picture, the spectra can be understood as a superposition of a SC background and ABS contributions. 
In PtBi$_2$, surface superconductivity has been reported to originate from Fermi-arc surface states with finite penetration into the bulk. 
Such Fermi-arc surface states are spin-momentum locked and connected to the underlying Weyl electronic structure in the bulk.
{This electronic structure may allow phase-sensitive scattering processes through which the phase structure of the SC order parameter influences the resulting ABS spectra.}
As illustrated in Fig.\,\ref{Figure5}g, such surface-extended ABSs can emerge at the interface between the surface SC region and the underlying bulk metallic states, with spectral weight extending towards the surface. 
The observed unit-cell-scale modulation of the in-gap spectral weight (Fig.\,\ref{Figure4}) is also consistent with this picture, as the corresponding quasiparticle wavefunctions can vary spatially within a unit cell and thereby give rise to a position-dependent tunneling response.
More generally, superconductivity-related Andreev spectral features can acquire unit-cell-scale spatial modulation through the interplay between Bloch wavefunctions, tunneling matrix elements, and the momentum-dependent SC order parameter \cite{AR_PRL_2023}. In the present case, such a mechanism may further contribute to the experimentally observed spatial variations of the in-gap spectral weight.
{Further theoretical work is in order to rationalise our observations beyond our empirical model.} 
We note that although STM/STS probes these quasiparticle states via single-particle tunneling, it does not provide direct access to their momentum-resolved dispersion. Consequently, the detailed band structure and topology of the ABS band cannot be unambiguously determined. Nevertheless, the systematic phenomenological simulation of the tunneling spectra may indicate a potential to chiral pairing symmetry in PtBi$_2$. 
\vspace{0.4cm}

{\section{D\MakeLowercase{iscussion}}}  
\label{Sec:Results} 
\vspace{-0.3cm}
	
{Recent ARPES studies on PtBi$_2$ have proposed an $i$-wave SC state, which we contrast here with the physical picture emerging from our STM/STS measurements \cite{iwave}. 
In Ref.~\cite{iwave}, a nodal SC gap of 1--2\,meV was reported on the Fermi-arc surface states, and the superconductivity was described as an anomalous 2D topological state that exists only as the boundary of a 3D crystal. Within this framework, surface Majorana-cone-like excitations and 1D Majorana flat bands at step edges were proposed. By contrast, our STM/STS measurements reveal a substantially larger gap $\sim$ 10\,meV. At the same time, the finite DOS observed within the gap prevents an unambiguous determination of whether this gap is fully open or nodal. In addition, ARPES has not yet resolved a gap of this magnitude on comparable spatial scales, so a direct comparison of the detailed gap structure remains inconclusive. Phenomenological modeling further shows that the experimentally observed spectral features are consistent with a chiral pairing scenario and the presence of a surface ABS band within the SC gap, indicating an unconventional SC state in PtBi$_2$. However, STM/STS measurements do not allow us to determine whether the ABS band corresponds to a surface Majorana-cone excitation. We therefore refrain from such an identification.}
\vspace{0.5cm}
	
\section{S\MakeLowercase{ummary}}
\label{subsec:Summary}
\vspace{-0.3cm}
{This work reports the experimental observation of a sizable gap and surface-extended in-gap states in PtBi$_2$. The possible physical origins of these features are systematically examined through additional control measurements combined with phenomenological modeling. We find that the experimental data are most consistently interpreted as surface-extended ABSs within a SC gap, and that their characteristic spectral features are best captured by a chiral pairing scenario. In this sense, our work provides direct real-space spectroscopic insight into the unconventional superconductivity in PtBi$_2$. These results establish PtBi$_2$ as a promising platform for exploring unconventional quasiparticle excitations and potential Majorana physics enabled by a sizable SC gap at elevated temperatures.}
\vspace{0.3cm}
	
\section{A\MakeLowercase{cknowledgments}}
\label{sect:Acknowledgments}
\vspace{-0.3cm}
{We thank Andrii Kuibarov and Sergey V. Borisenko for providing complementary ARPES data used in our response to the referees and for helpful discussions.}
X.H.~would like to thank for the financial support from the DFG 
through the Hallwachs-R{\"o}ntgen Postdoc Program of ct.qmat (EXC 2147, Project No. 390858490). 
L.Z.~would like to thank for the financial support from the Guangdong Provincial Quantum Science Strategic Initiative (GDZX2401011). 
M.B.\ acknowledges ct.qmat (EXC 2147, Project No. 390858490). 
C.H.\ and B.B.\ acknowledge support from the Deutsche Forschungsgemeinschaft (grant 566479091).

\section{A\MakeLowercase{uthor contributions}}
\label{sect:Acknowledgments}
\vspace{-0.3cm}
X.H., S.S., C.H., B.B., and M.B. conceived and supervised the project. 
L.Z. and {L.H.} synthesized the single crystals. X.H. carried out the STM/STS experiments with assistance from S.S. and P.H. 
The theoretical model was developed by S.S. and X.H. 
Under the supervision of C.H., S.S. and J.B. wrote the code and performed the simulations. 
X.H. wrote the initial draft of the manuscript. 
All the authors contributed to the interpretation of the results and to the revision of the manuscript. 


%

\clearpage

\section{M\MakeLowercase{ethods}}  
\label{Sec:M} 
\vspace{-0.3cm}

\noindent\textbf{The synthesis and characterization of PtBi$_2$ single crystals}

Single crystals of PtBi$_2$ were grown using the Bi-flux method \cite{R5}. Platinum (99.95\%) and bismuth (99.99999\%) grains were measured at a stoichiometric ratio 
of $\mathrm{Pt}:\mathrm{Bi} = 1:6$ and sealed into an evacuated quartz tube. 
The tube was heated to $800^{\circ}$C and maintained for 20 hours, and then cooled down to $450^{\circ}$C in 350 hours. 
After centrifugation, the single crystals were separated from the flux.
\medskip

\noindent\textbf{Preparation of atomically clean PtBi$_2$ surfaces}

\noindent Large, flat PtBi$_2$ single crystals, each with a lateral size exceeding ${(2 \times 2)}$ mm\textsuperscript{2}, 
were mounted on a molybdenum sample holder using silver epoxy and cleaved under UHV conditions using Kapton tape. 
\medskip

\noindent\textbf{Scanning tunneling microscopy/spectroscopy measurements}

The STM/STS measurements were performed at 5.1\,K 
using a UHV home-built low-temperature STM system ($p < 1.2 \times 10^{-10}$\,Torr). 
Electrochemically etched W-tips coated with silver were used for the measurements. 
Differential tunneling conductance (d$I$/d$V$) spectra were acquired using the lock-in technique 
in open feedback mode, with a modulation voltage ($V_{\text{mod}}$) at a frequency of 987.5\,Hz. 
STM image processing was performed using the WSxM software \cite{WSXM}.
\medskip

\noindent\textbf{Theoretical model}
\medskip

\noindent\textbf{{(1) Dynes model}}

{Within the Dynes model, the tunneling current {into a superconductor is described by the energy integrated angular average of the angle-dependent density of states  
$N_{\mathrm{SC}}(\varepsilon,\theta_k)$} \cite{Dynes1978}:}
\medskip
{
\begin{equation}
	{\begin{split}
	I_\mathrm{Dynes} ~& \propto \int_{-\infty}^{\infty} \frac{1}{2\pi}\int_0^{2\pi} N_{\mathrm{SC}}(\varepsilon,\theta_k) \\ & \left[f(\varepsilon, T) - f(\varepsilon - \mathrm{e}U, T)\right] \, \mathrm{d}\theta_k \, \mathrm{d}\varepsilon
    \end{split}
\label{Dynes}} 
\end{equation}
}
\noindent
{where $f(\varepsilon, T)$ is the Fermi--Dirac distribution function, {$\varepsilon$ is the energy relative to $E_{\rm F}$,} $U$ is the sample bias, $T$ is the temperature, and  $\theta_k$ denotes the polar angle in $k$-space. The angle-dependent density of states of a superconductor  
$N_{\mathrm{SC}}(\varepsilon,\theta_k)$ can be described by the Dynes formula:
\begin{equation}
	{N_{\mathrm{SC}}(\varepsilon,\theta_k) = \Re \left[\frac{\varepsilon - i \gamma}{\sqrt{(\varepsilon - i \gamma)^2 - |\Delta(\theta_k)|^2}}\right]
	\label{N_SC}}	
\end{equation}

\noindent
{with $\gamma$ being a phenomenological broadening parameter accounting for finite quasiparticle lifetime. 
In a general form, the momentum-dependent SC order parameter can be expressed as $\Delta_k \equiv \Delta(\theta_k) = \Delta_0 \cos(l\theta_k)e^{i\phi_k}$, where $\phi_k$ represents the momentum-dependent phase of the parameter, and the integer $l$ specifies the angular symmetry of the gap magnitude ($s$-wave: $l = 0$, $d$-wave: $l = 2$, ..., $i$-wave: $l = 6$). Accordingly, the gap magnitude entering Eq.\,\ref{N_SC} is $|\Delta(\theta_k)| = |\Delta_k| = \Delta_0 |\cos(l\theta_k)|.$
Note that, in the Dynes formulation adopted here, the absolute value of the order parameter is taken into account. Thus, consequences of a potential phase winding cannot be deduced.
\figEref{FigureE7}a,b show that in terms of the magnitude the Dynes model captures the qualitative difference between isotropic ($s$-wave or $p_x+ip_y$-wave) and anisotropic pairings (e.g. $i$-wave), giving rise to a U-shaped SC gap for isotropic pairing and a V-shaped SC gap for anisotropic pairing, respectively.
\medskip

\noindent\textbf{{(2) Blonder-Tinkham-Klapwijk (BTK) model}}
\medskip

{The BTK model is employed to describe Andreev reflection at the interface between a metallic STM tip and the SC surface of PtBi$_2$. The SC order parameter  $\Delta(\theta_k)$ is defined as described above, allowing for both isotropic and anisotropic gap structures. The tunneling current is expressed as \cite{BTK}:}
 
\begin{equation}
	{\begin{split}
		I_\mathrm{BTK} ~& \propto  \int_{-\infty}^{\infty} \int_0^{2\pi} N_{\mathrm{AR}}(\varepsilon,\theta_k) \\
		& \left[f(\varepsilon, T) - f(\varepsilon - \mathrm{e}U, T)\right] \, \mathrm{d}\theta_k \, \mathrm{d}\varepsilon
	\end{split}
	\label{BTK}} 
\end{equation}

\noindent {where $N_\mathrm{AR}$ describes the effective transmission coefficient including the SC quasiparticle density of states as well as contributions from Andreev reflections.}
\noindent {For $|\varepsilon| < |\Delta(\theta_k)|$, $N_\mathrm{AR}$ is given by:}
\begin{equation}
	{N_\mathrm{AR}(|\varepsilon| < |\Delta(\theta_k)|, Z) =  \frac{2|\Delta(\theta_k)|^2}{\varepsilon^2 + (|\Delta(\theta_k)|^2 - \varepsilon^2)(1 + 2Z^2)^2}
	\label{N_in}} 
\end{equation}
\noindent {For $|\varepsilon| > |\Delta(\theta_k)|$, $N_\mathrm{AR}$ is given by:}
\begin{equation}
	{N_\mathrm{AR}(|\varepsilon| > |\Delta(\theta_k)|, Z) = 1 + \frac{u_0^2 v_0^2}{\lambda^2} - \frac{v_0^2(u_0^2 -  v_0^2)^2 Z^2}{\lambda^2}
		\label{N_out}}  
\end{equation}

\noindent {with}
\begin{equation}
	{\lambda = u_\mathrm{0}^2 + Z^2(u_\mathrm{0}^2 - v_\mathrm{0}^2)
	\label{lambda}}	
\end{equation}

\noindent {and the BCS coherence factors:}

\begin{equation}
	{u_\mathrm{0}^2 = \frac{1}{2} (1 + \sqrt{\frac{\varepsilon^2 - |\Delta(\theta_k)|^2}{\varepsilon^2}})
	\label{u}}	
\end{equation}

\begin{equation}
	{v_\mathrm{0}^2 = 1 - u_\mathrm{0}^2
	\label{v}}
\end{equation}

{The dimensionless parameter $Z = k_F H/2E_F$ characterizes the effective interface barrier strength and is determined by the tip--sample junction. $H$ is the scattering potential at the junction, $k_F$ is the Fermi wave vector. In the simulation, a large $Z$ corresponds to the tunneling limit, yielding tunneling-dominated spectra, whereas decreasing $Z$ increases the interface transparency and drives a continuous evolution of the spectral line shape towards an Andreev-reflection-dominated regime.}
\vspace{0.3cm}

\noindent\textbf{{(3) Andreev bound state (ABS) model}}
\medskip

{The ABS current, $I_{\mathrm{ABS}}$, is modelled within the theoretical framework 
originally developed by Kashiwaya {\em et al.} \cite{ABS_Satoshi_2, ABS_Satoshi} and subsequently extended 
to topological superconductors and superfluid $^3$He \cite{ABS_3DTIS,R1}:}

\begin{equation}
	{\begin{split}
			I_\mathrm{ABS} ~& \propto \int_{-\infty}^{\infty} \int_0^{\pi/2} \Gamma_{\mathrm{ABS}}(\theta, \varepsilon) \\
			& \left[f(\varepsilon, T) - f(\varepsilon - \mathrm{e}U, T)\right] \,\mathrm{d}\theta \, \mathrm{d}\varepsilon
		\end{split}
		\label{ABS}} 
\end{equation}

{ 
\noindent In this phenomenological description, $\theta$ ($-\pi/2 < \theta < \pi/2$) denotes the quasiparticle incidence angle measured from the interface. Owing to the symmetry between the incident and reflected quasiparticle trajectories, it is sufficient to perform the angular integration over $0 < \theta < \pi/2$. The momentum component parallel to the interface is assumed to be conserved, and the angle-resolved ABS conductance kernel is written as:
\begin{equation}
{\begin{split}
		\Gamma_{\mathrm{ABS}}~&(\theta, \varepsilon) = \sin(\theta) \frac{\sigma_\mathrm{N}}{2} \\
		& \cdot  \sum\limits_{s = {\pm 1}} \frac{1 + \sigma_\mathrm{N} |N_\mathrm{ABS}(\varepsilon, \theta)|^2 
			+ (\sigma_\mathrm{N} - 1) |N_\mathrm{ABS}(\varepsilon, \theta)|^4}{\left|1 + (1 - \sigma_\mathrm{N}) N_\mathrm{ABS}(\varepsilon, \theta)^2 \cdot \exp(-i s \Phi)\right|^2}
			\end{split}}
		\label{Gamma_ABS}
\end{equation}
\noindent where $s = {\pm 1}$ labels the two quasiparticle trajectories connected by specular reflection at the interface. We note
that the phases of the SC order parameter associated with the reflected and incident trajectories are $\phi_{\mathbf{k}_r}$ and $\phi_{\mathbf{k}_i}$, respectively. Their difference, $\Phi = \phi_{\mathbf{k}_r} - \phi_{\mathbf{k}_i}$, enters the ABS contribution and is related to the incidence angle $\theta$ through the pairing state and the reflection geometry. For the chiral $p$-wave case considered here, this relation is $\Phi = 2\theta$.}	
The transmissivity at the interface is given by:

\begin{equation}
	{\sigma_\mathrm{N} = \frac{4 \cos^2(\theta)}{4 \cos^2(\theta) + Z_\mathrm{int}^2}}
\end{equation}

\noindent {where $Z_\mathrm{int} = k_{\rm F} H/2E_{\rm F}$.} {The auxiliary function entering the ABS conductance is defined as:} 

\begin{equation}
	{	N_\mathrm{ABS}(\varepsilon, \theta) = \frac{|\Delta(\theta)|}{\varepsilon + \sqrt{\varepsilon^2 - |\Delta(\theta)|^2}} }
\end{equation}
\noindent {Here, $|\Delta(\theta)|$ denotes the SC gap magnitude sampled by a quasiparticle trajectory with incidence angle $\theta$. In the phenomenological model employed here, its angular dependence is parametrized as
$|\Delta(\theta)| = \Delta_0 |\cos(l\theta)|$, where the integer $l$ specifies the angular symmetry of the gap magnitude.}

{Although the barrier parameters $Z$ in the BTK model and $Z_\mathrm{int}$ in the ABS model have the same mathematical form, they describe} physically distinct interfaces and should not be identified with each other. For the BTK description, $Z$ characterizes the effective tip--sample junction transparency and is therefore varied with tip--sample distance. By contrast, in the ABS model, $Z_\mathrm{int}$ describes the effective scattering strength at the intrinsic interface between the surface SC region and the underlying bulk metallic states, and therefore does not vary with the tip--sample distance.
Although changing $Z_\mathrm{int}$ can modify the detailed spectral line shape (see \figEref{FigureE7}c-f), $Z_\mathrm{int}$ is kept fixed in the simulating because this internal interface is an intrinsic property of the sample.
{We note that, within the ABS-based modeling employed here, the presence of an ABS-like band is effectively assumed at the SC-metal interface. This allows us to compare the spectral characteristics obtained for different phenomenological gap and phase structures within a unified framework. For the nonchiral isotropic reference case \figEref{FigureE7}c, we retain the ABS contribution and impose a fixed phase shift ($\Phi = \phi_{\mathbf{k}_r} - \phi_{\mathbf{k}_i} = \pi$)} in the Andreev scattering channel to emulate a bound-state-like condition. This construction serves as a reference case to compare the resulting spectral line shapes with those obtained for other pairing symmetries (\figEref{FigureE7}d-f). It is not intended to imply that a well-defined ABS band is expected for a generic isotropic, nonchiral SC-metal interface.
\vspace{0.2cm}

\noindent\textbf{{(4) Dynes + ABS model}}
\medskip

{Within the Dynes + ABS model, the tunneling current is written as:}

\begin{equation}
	{
		I = \left(1 - \mathrm{e}^{-\alpha}\right) \cdot I_{\mathrm{Dynes}}
		+ \mathrm{e}^{-\alpha} \cdot \, I_{\mathrm{ABS}}
	}
	\label{Dynes+ABS}
\end{equation}

{This formulation provides an effective description of combined Dynes and ABS contributions, with their relative weight controlled by the parameter $\alpha$. As $Z_\mathrm{int}$ is governed by the interface transparency of an intrinsic interface within the sample, it is taken to be constant ($Z_\mathrm{int} = 3$) for different tip--sample distances in this study. The experimentally observed evolution from SC-gap-dominated spectra to ABS-dominated spectra is therefore captured not by varying $Z_\mathrm{int}$, but by tuning $\alpha$. Figures\,\ref{Figure5}c-f present the simulated results using the combined Dynes + ABS model, while \figEref{FigureE7}g-j show calculations using the ABS model alone with $Z_\mathrm{int} = 3$ for clarity.}

The differential conductance shown in Fig.\,\ref{Figure5} and \figEref{FigureE7} is obtained as the derivative of the total tunneling current:
\begin{equation}
	\frac{dI}{dU} = D \cdot \frac{dI_{\mathrm{tot}}}{dU} + c
\end{equation}
The model output is scaled by $D$ and offset by $c$.


\setcounter{figure}{0}
\renewcommand{\thefigure}{E\arabic{figure}}
\renewcommand{\theHfigure}{E\arabic{figure}}

\begin{figure*}[t] 
	\centering
	\includegraphics[width=0.6\textwidth]{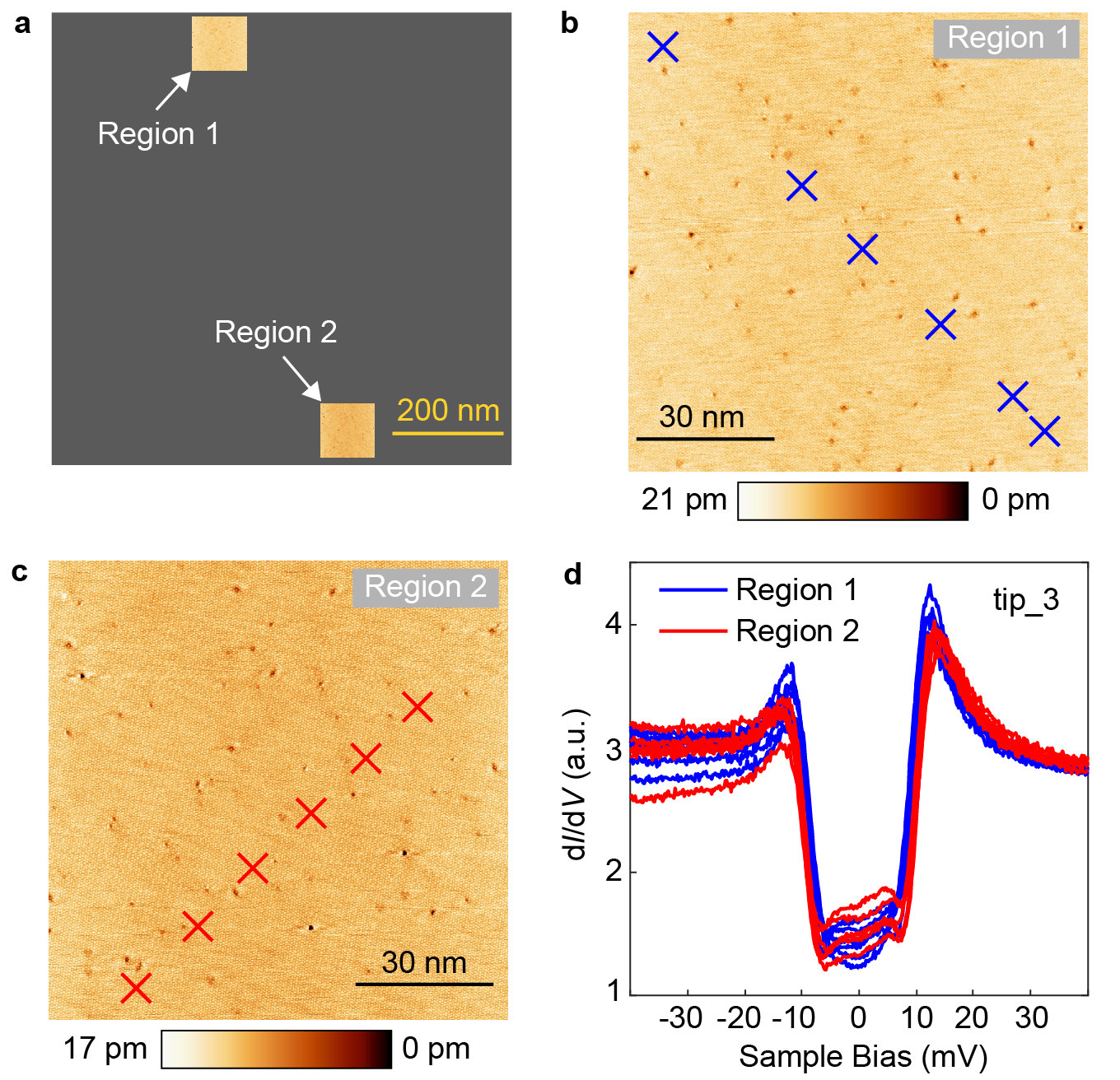}
	\caption{\textbf{Variations of {the sizable gap} over hundreds of nanometers in PtBi$_2$.} 
		\textbf{a}, Schematic illustration of the relative spacing between two measurement regions.
		\textbf{b}, STM image of region 1 ($V = 200$\,mV, $I = 50$\,pA). 
		\textbf{c}, STM image of region 2 ($V = 200$\,mV, $I = 100$\,pA). 
		The crosses in \textbf{b} and \textbf{c} mark the locations where the d$I$/d$V$ spectra were acquired.
		\textbf{d}, d$I$/d$V$ spectra measured (with tip\_3) at the marked positions in \textbf{b} and \textbf{c}. 
		Measurement conditions: $V_{\text{stab}} = 40$\,mV, $I_{\text{stab}} = 500$\,pA, and $V_{\text{mod}} = 0.5$\,mV. 
		$T$ = 5.1\,K. The gap sizes, determined from the coherence peak positions, range from 12.4\,meV, to 14.2\,meV, 
		with an average value of 13.05\,meV and a standard deviation of 0.80\,meV. All spectra are presented as raw data.}				 
	\label{FigureE1}
\end{figure*} 

\begin{figure*}[t] 
	\centering
	\includegraphics[width=0.65\textwidth]{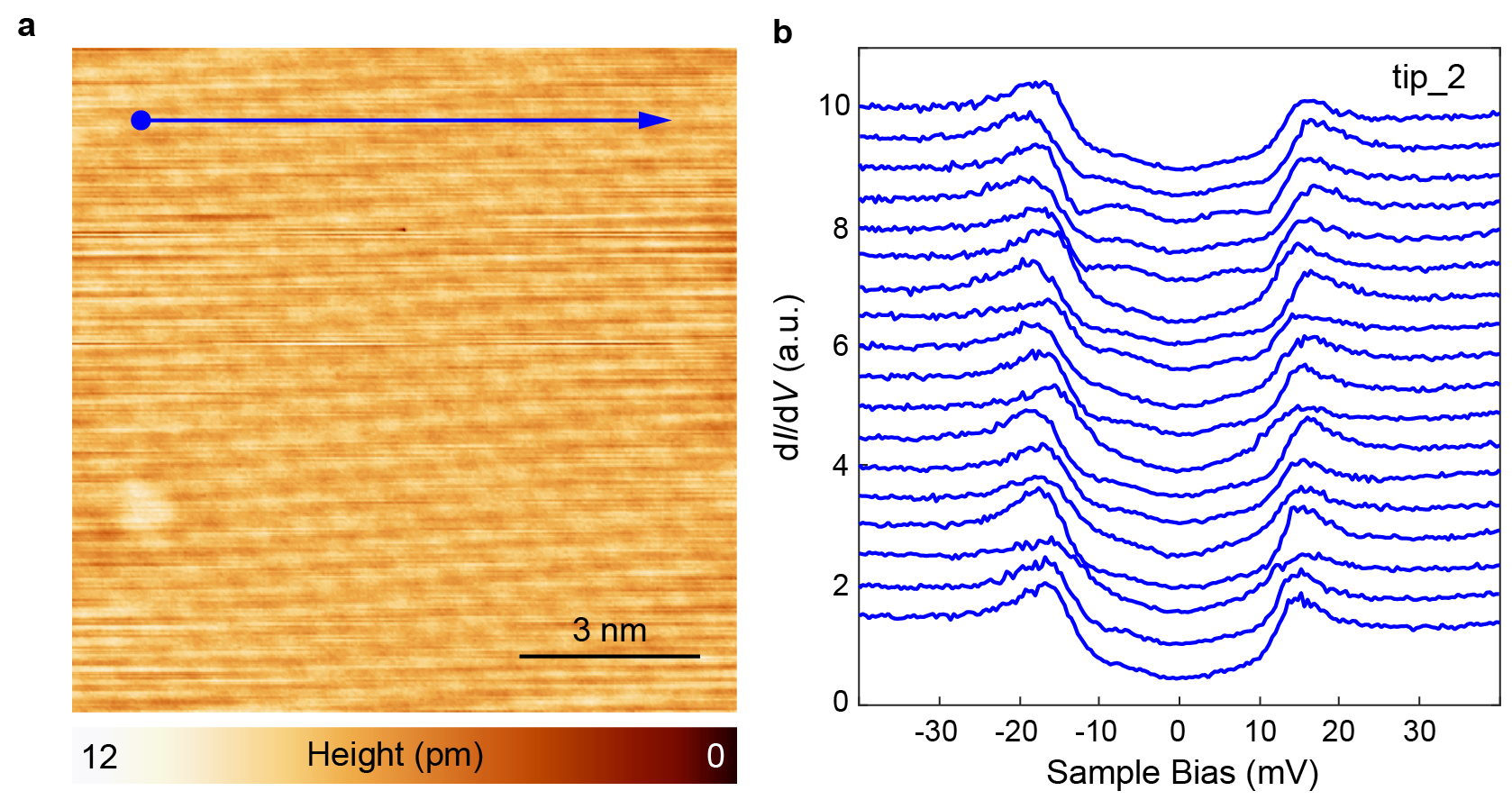}
	\caption{\textbf{Spatial homogeneity of {the sizable gap} in PtBi$_2$.} 
		\textbf{a}, Atomic-resolution STM image ($V = -200$\,mV, $I = 50$\,pA) of a ${(11 \times 11)}$ nm\textsuperscript{2} surface region.
		\textbf{b}, Spatially resolved d$I$/d$V$ spectra ($V_{\text{stab}} = 40$\,mV, $I_{\text{stab}} = 100$\,pA, 
		and $V_{\text{mod}} = 0.5$\,mV, tip\_2) acquired along the blue line indicated in a, 
		with a spacing of 0.5\,nm---smaller than the in-plane lattice constant of trigonal PtBi$_2$ ($a = b = 0.657$\,nm). 
		$T$ = 5.1\,K. Spectra are vertically offset for clarity. All spectra are presented as raw data.}				 
	\label{FigureE2}
\end{figure*} 

\begin{figure*}[t] 
	\centering
	\includegraphics[width=0.33\textwidth]{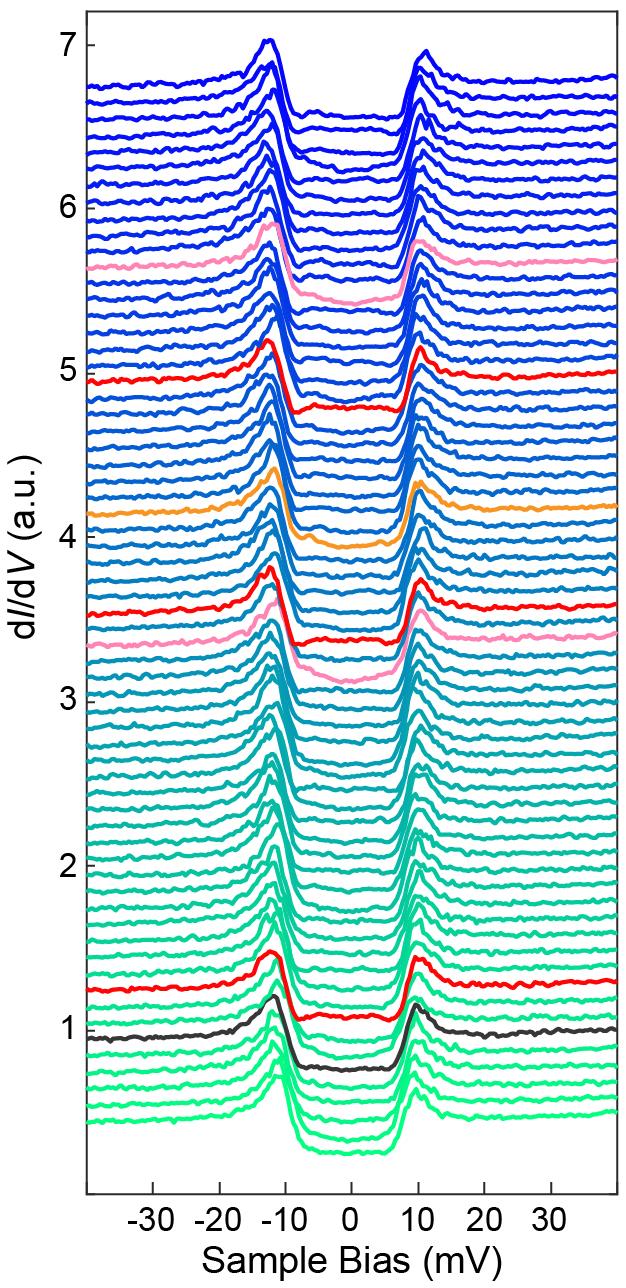}
	\caption{{\textbf{Same data set as in Fig.\,\ref{Figure2}b, with spectra vertically offset to emphasize features within {the gap}.}}
    {Some spectra exhibit a flat in-gap DOS (see curve colored black), 
	but others follow a convex (positive curvature, pink) or a concave profile (negative curvature, red).
	Of particular interest is the orange spectrum, which shows a parabolic-like DOS within {the gap}, 
	decorated by two small peaks symmetrically positioned around $E_{\text{F}}$.}}				 
	\label{FigureE3}
\end{figure*} 

\begin{figure*}[t] 
	\centering
	\includegraphics[width=0.95\textwidth]{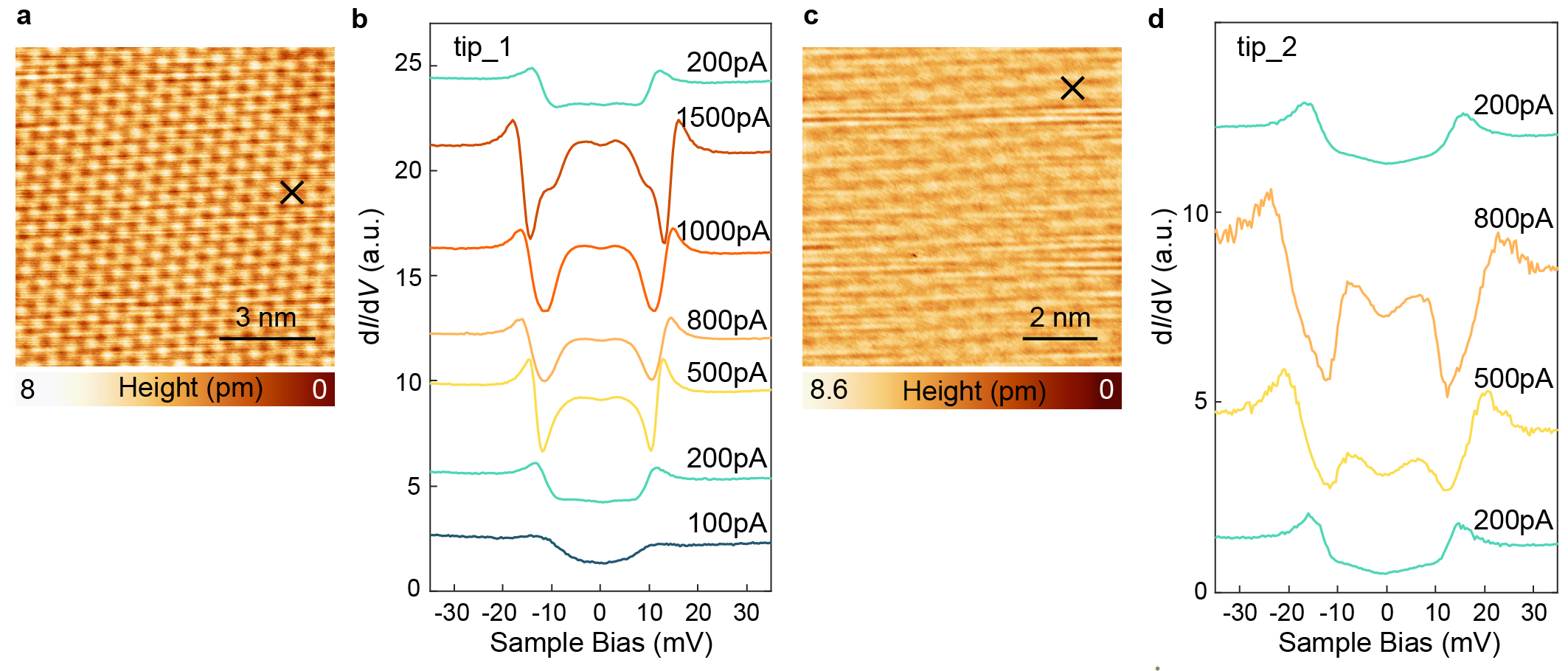}
	\caption{\textbf{Evolution of the in-gap states with tip-sample distance using different STM tips.} 
		\textbf{a}, Atomic-resolution STM image ($V = 200$\,mV, $I = 50$\,pA) 
			of a clean ${(10 \times 10)}$ nm\textsuperscript{2} region of PtBi$_2$.
		\textbf{b}, d$I$/d$V$ spectra acquired with tip\_1 at the position marked in \textbf{a}. 
			Measurement conditions: $V_{\text{stab}} = 40$\,mV, $V_{\text{mod}} = 0.5$\,mV. 
			The setpoint tunneling current was varied from 100\,pA to 1500\,pA and then reduced back to 200\,pA. 
		\textbf{c}, Atomic-resolution STM image ($V = -200$\,mV, $I = 50$\,pA) 
			of a clean ${(8.5 \times 8.5)}$ nm\textsuperscript{2} region of PtBi$_2$. 
		\textbf{d}, d$I$/d$V$ spectra acquired with tip\_2 at the position marked in \textbf{c}, 
			under a fixed sample bias of $V_{\text{stab}} = 40$\,mV. 
			The tunneling current was varied from 200\,pA to 800\,pA and then returned to 200\,pA. 
			$V_{\text{mod}} = 0.5$\,mV. $T$ = 5.1\,K. 
			Spectra in \textbf{b} and \textbf{d} are vertically offset for clarity. All spectra are presented as raw data.}				 
	\label{FigureE4}
\end{figure*} 

\begin{figure*}[p] 
	\centering
	\includegraphics[width=0.95\textwidth]{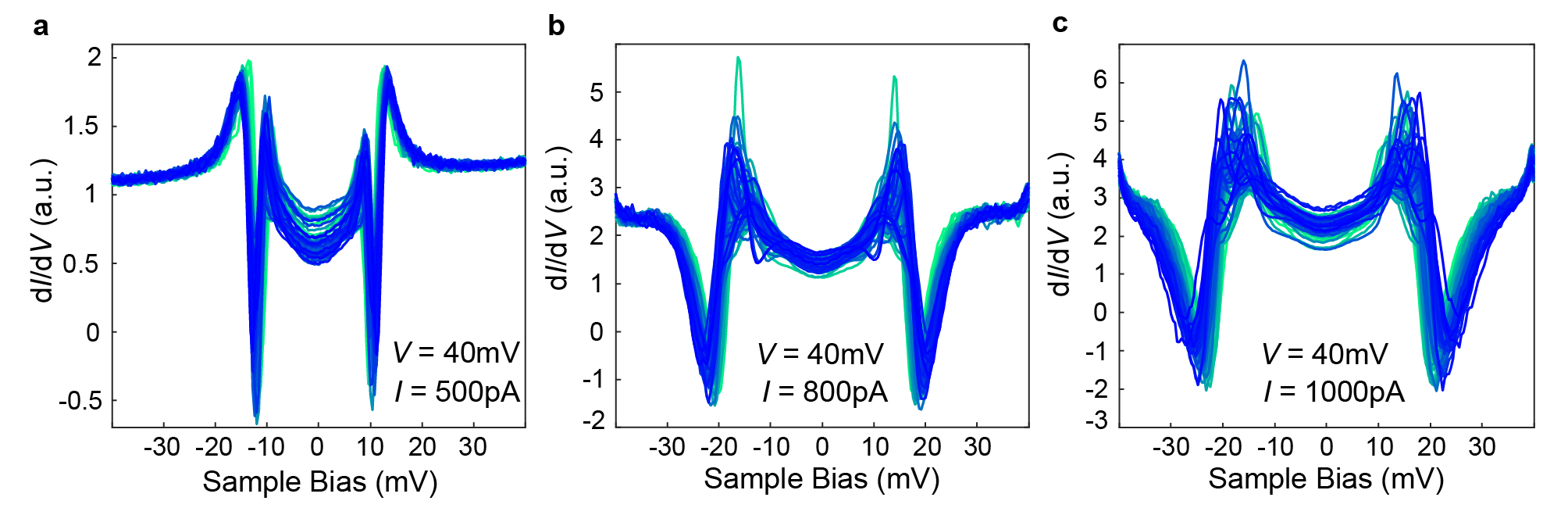}
	\caption{\textbf{Overlaid d$I$/d$V$ spectra from the same data sets shown in Fig.\,\ref{Figure3}}. 
		Highlighting the consistent spectral evolution across the measurement grid 
		and the emergence of negative differential conductance. All spectra are presented as raw data.}				 
	\label{FigureE5}
\end{figure*} 

\begin{figure*}[p] 
	\centering
	\includegraphics[width=0.55\textwidth]{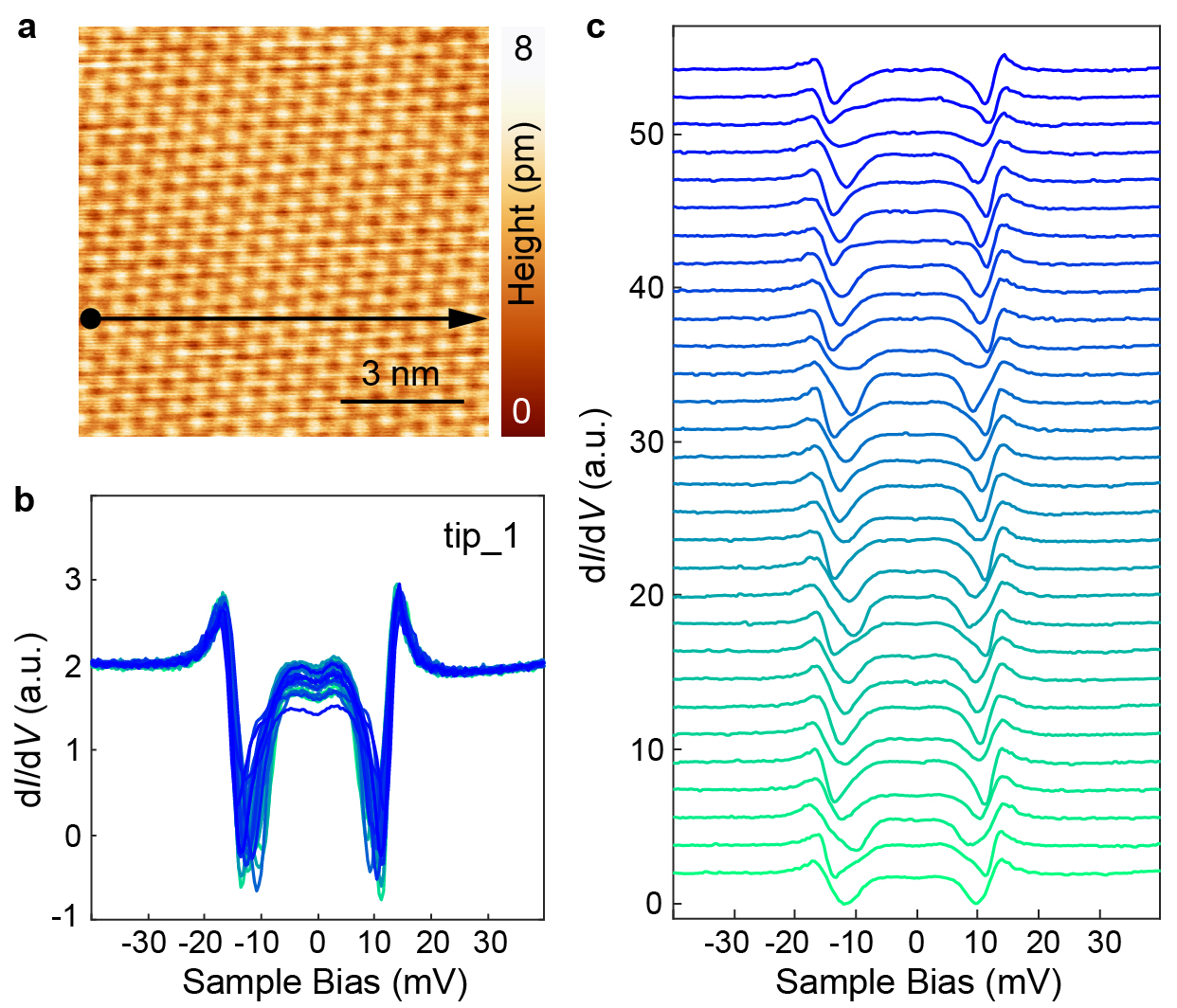}
	\caption{\textbf{High-resolution measurements of spatially extended {in-gap states} using tip\_1}. 
	\textbf{a}, Atomic-resolution STM image ($V = 200$\,mV, $I = 50$\,pA) 
		of a clean ${(10 \times 10)}$ nm\textsuperscript{2} region on the PtBi$_2$ surface.
	\textbf{b}, Spatially resolved d$I$/d$V$ spectra ($V_{\text{stab}} = 40$\,mV, $I_{\text{stab}} = 800$\,pA, 
		and $V_{\text{mod}} = 0.5$\,mV, tip\_1) acquired along the black line indicated in \textbf{a}, 
		with a spacing of 0.333\,nm. $T$ = 5.1\,K. 
	\textbf{c}, Same data as in \textbf{b}, shown with a vertical offset. All spectra are presented as raw data.}				 
	\label{FigureE6}
\end{figure*} 

\begin{figure*}[p] 
	\centering
	\includegraphics[width=0.95\textwidth]{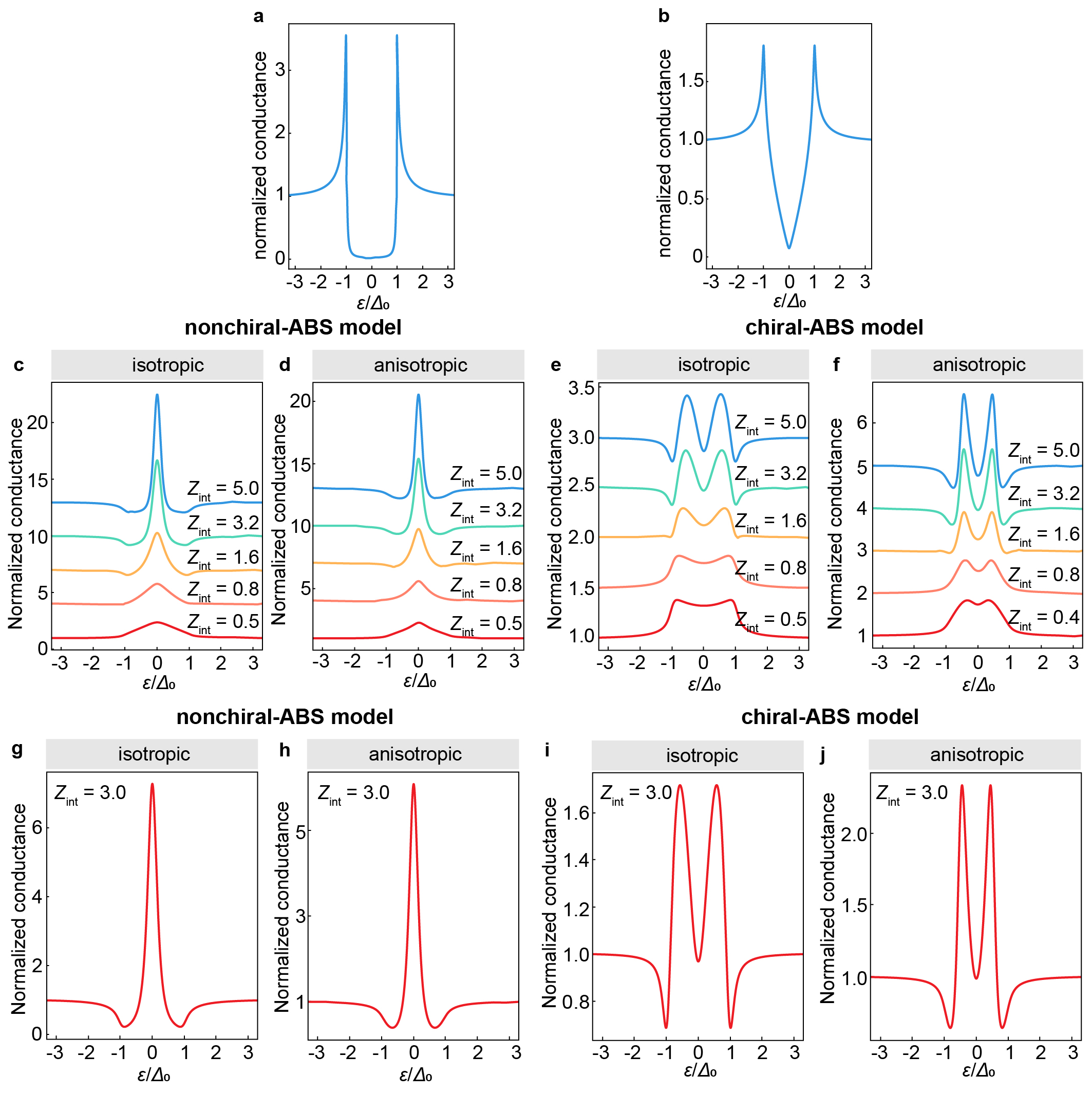}
	\caption{\textbf{{Simulated conductance based on different theoretical models}}. 
		{\textbf{a}, Simulated conductance within the isotropic Dynes model.	
		\textbf{b}, Simulated conductance within the anisotropic Dynes model.
		\textbf{c,d}, Simulated conductance within the nonchiral-ABS model for different $Z_\mathrm{int}$. \textbf{c}: isotropic pairing, \textbf{d}: anisotropic pairing.
		\textbf{e,f}, Simulated conductance within the chiral-ABS model for different $Z_\mathrm{int}$. \textbf{e}: isotropic pairing, \textbf{f}: anisotropic pairing. Spectra in \textbf{c-f} are vertically offset for clarity.
		\textbf{g,h}, Simulated conductance within the nonchiral-ABS model with $Z_\mathrm{int} = 3$. \textbf{g}: isotropic pairing, \textbf{h}: anisotropic pairing.
		\textbf{i,j}, Simulated conductance within the chiral-ABS model with $Z_\mathrm{int} = 3$. \textbf{i}: isotropic pairing, \textbf{j}: anisotropic pairing.}}				 
	\label{FigureE7}
\end{figure*} 

\end{document}